\documentstyle[aps,floats,psfig,epsf,twocolumn]{revtex}
\draft

\begin{document}
\title{Antiferromagnetic fluctuations, symmetry and shape of the gap function
\\in the electron-doped superconductors: the functional renormalization-group analysis}
\author{A. A. Katanin$^{a,b}$ }
\address{$^a$Max-Planck-Institut f\"ur Festk\"orperforschung, D-70569 Stuttgart,
Germany\\
$^b$Institute of Metal Physics, 620219 Ekaterinburg, Russia}

\address{~\\
\parbox{14cm}{\rm \medskip \vskip0.2cm
The problem of the symmetry of the superconducting
pairing and the form of the gap function in the electron-doped
superconductors is reconsidered within the temperature-cutoff functional
renormalization group approach combined with the Bethe-Salpeter equations.
The momentum dependence of the order parameter for antiferromagnetic and superconducting
instabilities in these compounds is analyzed. The gap function in the
antiferromagnetic (particle-hole) channel has its maxima at the hot-spots,
or at the diagonal of the Brilloin zone in their absence. The wavefunction in the singlet
superconducting channel is non-monotonic in the vicinity of the ($\pi ,0$)
and ($0,\pi $) points, {deviating therefore from the conventional d-wave
form} in striking similarity with recent experimental data. An
instability in the triplet superconducting channel is much weaker than the
singlet one and has an f-wave like form of the gap function.
}}

\maketitle

\tighten

It is by now well established that the superconducting order parameter in
high-T$_c$ compounds is described by a nearly ($\cos k_x-\cos k_y$)-momentum
dependence: its absolute value is largest at the Fermi surface (FS) points close to $(\pi ,0)
$ and $(0,\pi )$ and vanishes at the FS crossings on the Brillouin zone (BZ)
diagonals.

Accurate measurements of the gap function in the hole-doped
cuprates found, however, a slight deviation from this d$_{x^2-y^2}$-wave
momentum dependence \cite{Mesot}, with a flatter angular dependence near the
nodal points. On the other hand, recent experiments on the electron-doped cuprates
revealed non-monotonicity of the gap function as a function of an angle
around the FS \cite{Bluemberg,Matsui}. In particular, the observed gap function has its maxima away from the points
of the Fermi surface which are closest to the $(\pi,0)$ and $(0,\pi)$ points of the
Brillouin zone. Since the symmetry and the details of the momentum dependence of the
superconducting order parameter are closely related to the structure of the
effective pairing interaction between the electrons, the momentum dependence of the gap function contains
important information about the pairing mechanism in high-$T_c$ compounds.

One of the proposed candidates for the pairing mechanism in cuprates is the antiferromagnetic (AFM) spin fluctuations scenario \cite{Scalapino,BSW,Pines,Chubukov}. While the theoretical
studies of the hole-doped compounds found that antiferromagnetic
fluctuations lead to a d-wave symmetry of the gap function \cite{SLH,SWhte},
there is currently a theoretical controversy about the symmetry of the
superconducting pairing in the electron-doped cuprates. As it was argued earlier \cite{Bluemberg,Matsui},
with increasing electron doping the hot-spots of the Fermi surface (i.e. the points connected by the
antiferromagnetic wave vector $\bf{Q}=(\pi ,\pi )$) move towards each other (see Fig.1a). Since the sign of
the d-wave gap is opposite at the hot-spots, this leads to the suppression of the tendency towards the d-wave pairing.

In this situation, different
types of the pairing symmetry were proposed. An s-wave pairing at large electron dopings
was proposed\cite{Abrikosov}, as arising mainly due to the effect of disorder in the model with both short- and
long-range Coulomb parts of the interaction. On the other hand, for short-range interactions in the
absence of disorder, p-wave superconductivity was considered as a possible candidate
in the presence of strong charge fluctuations \cite{Yakovenko}. With these theoretical proposals,
it remains an important problem which pairing symmetry is favored at intermediate and large electron
dopings and what is the theoretically expected
ground state superconducting order parameter for the electron-doped superconductors.

To study this problem, we apply in the present paper the functional renormalization-group (fRG) technique \cite
{Zanchi,Metzner,SalmHon,SalmHon1,KK}, which proved successful in describing the interplay of antiferromagnetism and
d-wave superconductivity in the $t$-$t^{\prime }$ Hubbard model. Although a previous application of this technique
to the microscopic analysis of electron-doped superconductors within the Hubbard model found d-wave pairing even at relatively
high dopings \cite{Carsten}, the results of this study will be reconsidered in the present paper in two points.

First, the momentum-cutoff renormalization group approach of Ref. \cite{Carsten} does not
allow to search for the charge and spin instabilities in the forward scattering
(zero-momentum transfer) channel. This approach,
therefore, may miss the possibility of the triplet pairing, which is often closely related to such instabilities.
This drawback is overcome in the
temperature-cutoff fRG approach \cite{SalmHon1}, which uses temperature as a natural cutoff parameter
and proved successful in describing both AFM and ferromagnetic instabilities together with
singlet- and triplet superconducting pairing \cite{SalmHon1,KK}.

Second, a recent extension of the temperature cutoff renormalization group approach - its combination with the
Bethe-Salpeter (BS) approach - was proposed as a very
convenient tool of studing symmetry and the form of the wavefunctions of different order parameters \cite{BS}.
Contrary to previous fRG approaches, this method is not based on the knowledge of the wavefunctions at high temperatures.
For the hole-doped superconductors
it predicts correctly the flattening of the gap function near the nodal points \cite{BS} in agreement with
the experimental observation \cite{Mesot}.

In the present paper we use the temperature cutoff functional renormalization group approach in combination
with the Bethe-Salpeter equations to extract eigenfunctions and eigenvalues of the effective interaction in the
particle-particle (pp) or the particle-hole (ph) channel in the
electron-doped superconductors.

\begin{figure}[t!]
\psfig{file=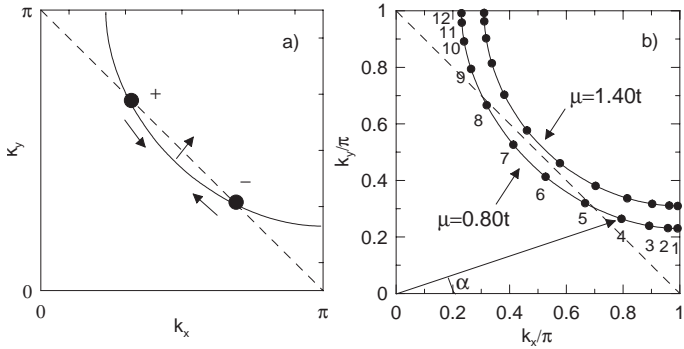,width=90mm,silent=} \vspace{2mm}
\caption{(a) The shape of the Fermi-surface (solid line) in the electron-doped
superconductors, the dashed line shows the umklapp surface $k_x+k_y=\pi $, black
circles mark hot-spots, arrows indicate the direction of movement of the Fermi-surface
and hot-spots with increasing doping. (b) The actual non-interacting Fermi surfaces
of the model (\ref{H}) at $t^{\prime }=0.3t,$ and $\mu =0.8t$ ($\delta=n-1=0.09$)
and $\mu =1.4t$ ($\delta=0.26$). The dots mark the positions of the centers of 12 patches,
located in the upper-right quarter of the Brilloin zone; $\alpha $ is an angle
corrdinate of the patch centers.}
\label{fig:Fig1}
\end{figure}

{\it The model}. As a convenient model which describes electron-doped superconductors, we
consider the 2D $t$-$t^{\prime }$ Hubbard model $H_\mu =H-(\mu -4t^{\prime
})N$ with
\begin{equation}
H=-\sum_{ij\sigma }t_{ij}c_{i\sigma }^{\dagger }c_{j\sigma
}+U\sum_in_{i\uparrow }n_{i\downarrow }  \label{H}
\end{equation}
where $t_{ij}=t$ for nearest neighbor (nn) sites $i$ and $j$ and $%
t_{ij}=-t^{\prime }$ for next-nn sites ($t,t^{\prime }>0$) on a square
lattice. The chemical potential $\mu$ is determined by the filling
$n>1$; for convenience we have shifted the chemical potential $\mu $ by $%
4t^{\prime }$.

{\it The method}. We follow the many-patch temperature cutoff fRG for one-particle
irreducible Green functions as proposed in Ref. \cite{SalmHon1}. This scheme accounts for
excitations with momenta far from and close to the FS, which is necessary
for the description of instabilities arising from zero-momentum transfer ph
scattering. The momentum dependence of the effective interaction
between electrons at a given temperature, $V_{T}({\bf k}_1,{\bf k}_2,{\bf k}_3,{\bf k}_4)$ is parametrized
by the position of incoming ${\bf k}_1,{\bf k}_2$ and outgoing electron momentum
${\bf k}_3$ at the Fermi surface. The fourth momentum, ${\bf k}_4$ is determined by the
momentum conservation law. Neglecting the frequency dependence of the
vertices, which is expected to have minor relevance in the weak-coupling
regime, the RG differential equation for the interaction vertex has the form
\cite{SalmHon1}
\begin{eqnarray}
\frac{{\rm d}V_T}{{\rm d}T}=-V_T\circ \frac{{\rm d}L_{{\rm pp}}}{{\rm d}T}%
\circ V_T+V_T\circ \frac{{\rm d}L_{{\rm ph}}}{{\rm d}T}\circ V_T\,,
\label{dV}
\end{eqnarray}
where $\circ $ is a short notation for summations over intermediate momenta
and spin,
\begin{equation}
L_{\text{ph,pp}}({\bf k},{\bf k}^{\prime })=\frac{f_T(\varepsilon _{{\bf k}%
})-f_T(\pm \varepsilon _{{\bf k}^{\prime }})}{\varepsilon _{{\bf k}}\mp
\varepsilon _{{\bf k}^{\prime }}},  \label{Lphpp}
\end{equation}
and $f_T(\varepsilon )$ is the Fermi function. The upper sign in Eq.(\ref
{Lphpp}) is for $L_{\text{ph}}$ and the lower sign for $L_{\text{pp}}$, respectively. Eq.(%
\ref{dV}) has to be solved with the initial condition $V_{T_0}({\bf k}_1,%
{\bf k}_2,{\bf k}_3,{\bf k}_4)=U$; the initial temperature is chosen as
large as $T_0=400t$.

We discretize the momentum space in $N_p=48$ patches using the same patching
scheme as in Ref. \cite{SalmHon1}. This reduces the integro-differential
equations (\ref{dV}) to a set of 5824 differential equations,
which were solved numerically. The position of the centers of the patches at the
Fermi surface for two different filling is shown in Fig. 1b.

To perform an analysis of possible instabilities within the fRG+BS approach, we
consider the solution of the Bethe-Salpeter equations \cite{BS,BetheSal}
\begin{eqnarray}
\sum_{{\bf p}}\Gamma _{\text{ph}}^T({\bf k},{\bf p})L_{\text{ph}}(%
{\bf p},{\bf p}+{\bf Q})\phi _{{\bf p}}^{\text{ph}} &=\displaystyle{\frac{\lambda _{\text{ph}}\phi _{%
{\bf k}}^{\text{ph}}}{1-\lambda _{\text{ph}}}}
\,,  \nonumber \\
-\sum_{{\bf p}}\Gamma _{\text{pp}}^T({\bf k},{\bf p}%
)L_{\text{pp}}({\bf p},-{\bf p})\phi _{{\bf p}}^{\text{pp}} &=\displaystyle{\frac{\lambda _{\text{pp}}\phi _{%
{\bf k}}^{\text{pp}}}{1-\lambda _{\text{pp}}}}\,,  \label{BSirr}
\end{eqnarray}
The 2-particle reducible vertices $\Gamma _{\text{ph,pp}}^T({\bf k},{\bf k}^{\prime })$ can be directly extracted
from the fRG flow according to

\begin{equation}
\Gamma _{\text{ph,pp}}^T ({\bf k},{\bf k}^{\prime })=\left\{
\begin{array}{cl}
V_T({\bf k},{\bf k}^{\prime },{\bf k}^{\prime }+{\bf Q}) \text{,\ \ \ ph (AFM)} \\
V_T({\bf k},-{\bf k},{\bf k}^{\prime })\pm V_T({\bf k},-{\bf k},-{\bf k}^{\prime }), \\
\text{\ \ \ \ \ \ \ \ \ \ \ \ pp (singlet, triplet SC)}
\end{array}
\right.   \label{Gamma}
\end{equation}

The value $\lambda _{\text{ph,pp}}=1$ corresponds to an ordering instability with the
symmetry of the eigenfunction $\phi _{{\bf k}}^{\text{ph,pp}}$. Therefore, tracing
the temperature dependence of eigenvalues and -functions allows to identify
both, the leading instabilities {\it and} their concomitant order parameter
structure. We stop the fRG flow at the temperature $T_X=0.001t$ (the maximum
interaction vertex $V_{\max}\equiv\max\{V({\bf k}_1,{\bf k}_2;{\bf k}_3,
{\bf k}_4)\}$ at this temperature remains smaller than the bandwidth).
We have verified that the results for the eigenfunctions are only weakly dependent
on the choice of $T_X$.

\begin{figure}[t!]
\psfig{file=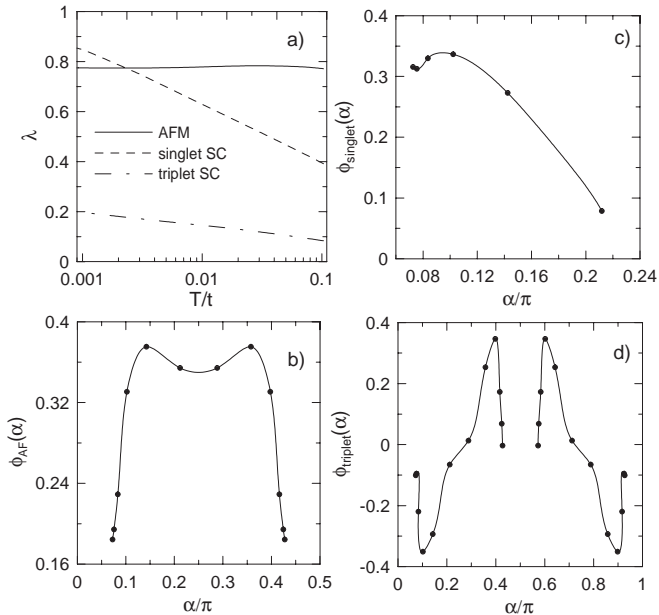,width=90mm,silent=} \vspace{2mm}
\caption{Eigenvalues (a) and angular-dependence on the
FS of the eigenfunctions $\phi _{{\bf p}}$ of the
Bethe-Salpeter equation at $T=T_X$ (b,c,d) for
$t^{\prime }=0.3t,$ $U=3.5t,$ $\mu=0.8t$ ($\delta=0.09$).
$T_X$ is the lowest temperature reached in the fRG flow.
The dots are the value of the eigenfunctions at the centers of corresponding
patches, cf. Fig. 1}
\label{fig:Fig2}
\end{figure}

{\it Results.} Below we discuss the results of the numerical solution of the
Bethe-Salpeter equations in the electron-doping regime. We choose $t^{\prime
}=0.3t$ which is close to typical values considered for the electron-
doped cuprates. To remain in the weak-coupling regime, where the
considered technique is applicable, we put $U=3.5t.$ Although this value is
substantially smaller than that which is expected for the cuprate materials, we show
that it allows us to reproduce the main features which are observed experimentally.
We also do not expect qualitative change of the results even for higher values of $U.$


We start in Fig. 2 with the results for the chemical potential $\mu
=0.8t,$ which corresponds to a filling $n=$ $1.09,$ i.e. $9\%$ of the
electron doping. At this filling the eigenvalue corresponding to the AFM
instability saturates at a value $\lambda<1$ at the smallest temperature $T_X$
which one can reach in the fRG flow. Therefore, the AFM is
not expected to be the leading ground-state instability
(more generally, the AFM is found not to dominate at $\mu>0.7t$, i.e. $\delta>0.05$). The
corresponding eigenfunction in the particle-hole channel (Fig. 2b) has its maxima at the hot-spots of the
Fermi surface.

The eigenvalue $\lambda_{pp}$ in the singlet pairing channel is monotonically increasing with decreasing
temperature, and it is
biggest close to $T_X$. Therefore, the singlet pairing
is expected to be the leading instability at $T\rightarrow 0$. The corresponding wave function
is non-monotonic (Fig. 2c) and has a shape which is strikingly
similar to the recent experimental data \cite{Bluemberg,Matsui}.
The maximum of the wavefunction lies between the 3-rd and 4-th patch of the Fermi surface,
i.e. it is shifted from the position of the hot-spot towards
the point of the FS closest to the $(\pi ,0)$. The eigenvalue corresponding
to the triplet pairing instability is much smaller than for the singlet one,
although it increases with decreasing temperature.
The corresponding wave function (Fig. 2d) has nodes at the diagonals, and it is maximal near hot spots
having therefore f-wave like symmetry.

\begin{figure}[t!]
\psfig{file=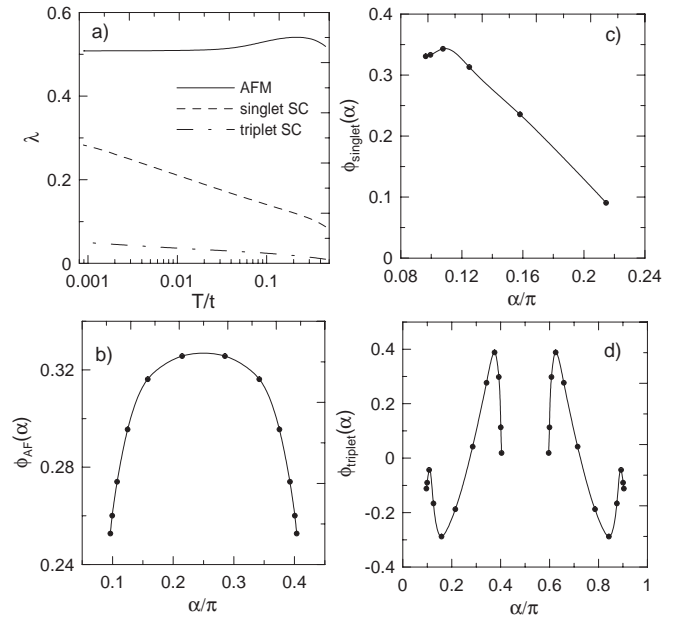,width=90mm,silent=} \vspace{2mm}
\caption{Same as Fig. 2 for $\mu=1.4t$ ($\delta=0.26$).}
\label{fig:Fig3}
\end{figure}

Now we consider the case of stronger doping $\mu =1.40t,$ i.e. $%
n=1.26$ (26\% of the electron doping), see Fig. 3. In this case the hot-spots of the Fermi surface
are absent. The eigenvalue corresponding to the AFM instability is decreasing
with decreasing temperature at low $T$, so that this instability is again not favored in the ground state.
The corresponding eigenfunction has its maximum at the diagonal of the Brillouin zone.
Both eigenvalues corresponding to the singlet- and triplet
superconducting pairing are smaller than in the case of $\mu =0.80t,$ although the
singlet superconductivity remains the leading instability in the $T\rightarrow 0$ limit.
The shapes of the wavefunctions
for singlet and triplet pairing are similar to those at $\mu =0.80t$.
The maximum of the wavefunction in the singlet superconducting channel is further shifted
towards the $(\pi ,0)$ point of the Brillouin zone (i.e. in the direction opposite
to the hot-spots), while maxima of the eigenfunction in the triplet pairing
channel are shifted towards the diagonal.

Now we discuss the physical origin of the shape of the gap functions found in the numerical
analysis. To this end, we plot the effective pairing interaction $\Gamma _{pp}^T ({\bf k},{\bf k}^{\prime })$
in the singlet- and triplet channel as a function of the momenta ${\bf k},{\bf k}^{\prime }$ on
the Fermi surface (Fig.4). One can see that at low electron doping ($\mu=0.8t$) the maxima
of the attractive interaction in both, the singlet and the triplet channels are located near the hot-spots (Fig. 4a,b), which leads
to the maximum of the pairing gap near these points. The difference in 1 patch between the position of the hot-spots
and the maxima of the gaps can be explained in this case by small incommensurability of spin fluctuations and the contribution of
the other channels of electronic scattering. With increasing doping the maximum of the interaction in the singlet channel spreads in a broader momentum range
(Fig. 4c), which, however, leads to almost the same position of the maxima of the gap, as for small electron doping.
The maximum of the attraction in the triplet channel at the points with
${\bf k}_{x,y}=-{\bf k}_{y,x}^{\prime }$ shifts
towards the diagonal of the Brillouin zone, where it is compensated, however, by the strong repulsion, which arises at ${\bf k}={\bf k^{\prime}}$ (Fig. 4d).
As a ``compromise", the maxima of the gap are located in the parts of the momentum space where neither repulsive nor attractive
interaction is strong, i.e. again remain almost unchanged with respect to small doping.

\begin{figure}[t!]
\psfig{file=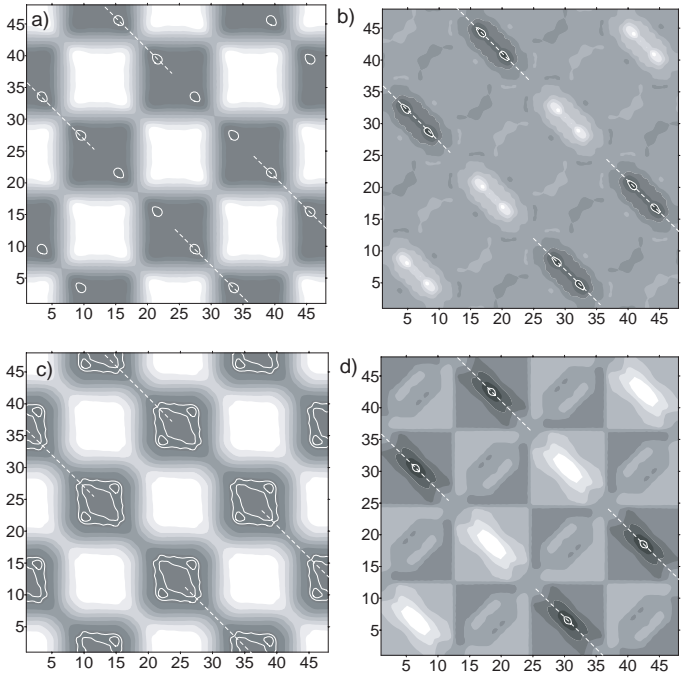,width=90mm,silent=} \vspace{2mm}
\caption{Contour plots of the pairing interaction $\Gamma _{pp}^T ({\bf k},{\bf k}^{\prime })$
in singlet (a,c) and
triplet (b,d) channels at $t^{\prime }=0.3t,$ $U=3.5t,$ $\mu=0.8t$ (a,b)
and $\mu=1.4t$ (c,d) as a function of the position of the momenta ${\bf k},{\bf k}^{\prime }$ at the Fermi surface
(labeled by the number of patch). Black color corresponds to the attractive (negative),
white - to the repulsive (positive) interaction. White contours show regions
with the strongest attraction, dashed lines mark the points with ${\bf k}_{x,y}=-{\bf k}_{y,x}^{\prime }$.
}
\label{fig:Fig4}
\end{figure}

Therefore, while at small doping the shape of the gap functions are determined by the spin fluctuations
with the wavevector close to $\bf Q$, at larger dopings the spin fluctuations with broader range of momenta
start to play an important role. Note that the maximum of the singlet gap is
located away from the Fermi surface points closest to $(\pi ,0)$ and $(0, \pi )$ only for the fillings $n>1$;
on the hole doped side (i.e. at fillings $n<1$) the nonmonotonicity of the singlet gap function quickly disappears \cite{BS}.

In conclusion, we have investigated the symmetry of the leading
instabilities and the shape of the corresponding wavefunctions of the
2D electron doped superconductors within the 2D $t$-$t^{\prime }$
Hubbard model using as a novel tool the combination of the Bethe-Salpeter
equation and the fRG approach. At $U=3.5t$ we have found the $d$-wave pairing
instability to be the strongest instability for dopings $\delta > 0.05$
(one can expect that the range of dopings where the AFM instability is the leading one,
increases further with increasing interaction strength). The angular dependence of the singlet pairing
gap function is found to be non-monotonic and its shape is strikingly similar to the
recent experimental data. The position of the maximum of the gap is close to the position
of the hot spots at low doping level and deviates from it at higher dopings, where
the antiferromagnetic fluctuations with wavevectors with broad momenta range become
essential. The triplet instability has subleading eigenvalues and f-wave like
form of the wavefunction. The maxima of the wavefunction are close to the hot-spots
at low electron dopings and are close to the diagonal for higher dopings.

I am grateful to V. M. Yakovenko for stimulating discussions and pointing my attention
to the problem of electron-doped cuprates and to the ICCMP (Brasilia) where
this work was initiated.

\end{document}